# A collaborative framework to exchange and share product information within a supply chain context

H.M. Geryville[a,b], Y. Ouzrout[a], A. Bouras[a], N.S. Sapidis[b]

[a] PRISMa-Lyon2 laboratory, IUT Lumière, 160 boulevard de l'université 69676 Bron Cedex France
hichem.geryville@univ-lyon2.fr, yacine.ouzrout@univ-lyon2.fr, abdelaziz.bouras@univ-lyon2.fr

[b] Department of Product and Systems Design Engineering, University of the Aegean, Ermoupolis–Syros 84100 Greece
sapidis@syros.aegean.gr

*Abstract*— The new requirement for "collaboration" between multidisciplinary collaborators induces to exchange and share adequate information on the product, processes throughout the products' lifecycle. Thus, effective capture of information, and also its extraction, recording, exchange, sharing, and reuse become increasingly critical. These lead companies to adopt new improved methodologies in managing the exchange and sharing of information. The aim of this paper is to describe a collaborative framework system to exchange and share information, which is based on: (i) The Product/Process/Collaboration/Organization model (PPCO) which defines product and process information, and the various collaboration methods for the organizations involved in the supply chain. (ii) Viewpoint model describes relationships between each actor and the comprehensive Product/Process model, defining each actor's "domain of interest" within the evolving product definition. (iii) A layer which defines the comprehensive organization and collaboration relationships between the actors within the supply chain. (iv) Based on the above relationships, the last layer proposes a typology of exchanged messages. A communication method, based on XML, is developed that supports optimal exchange/sharing of information.

To illustrate the proposed framework system, an example is presented related to collaborative design of a new piston for an automotive engine. The focus is on user-viewpoint integration to ensure that the adequate information is retrieved from the PPCO.

*Keywords*— Product Lifecycle Management (PLM), Supply Chain Management (SCM), Collaboration, Viewpoint, Product Information Exchange.

## I. INTRODUCTION

As companies are moving towards providing better and better customer-centric products and services to improve market share and market size with continuously growing revenue, the efficiency and effectiveness of product lifecycle management becomes much more important. Product Lifecycle Management (PLM) is a strategic business approach that applies a consistent set of business solutions to support the collaborative creation, management, dissemination, and use of product definition information across the extended enterprise from concept to end of life – integrating people, processes, and information [1]. Moreover, for more efficiency these companies need to integrate their business partners in the collaboration to optimize PLM processes by integrating some Supply Chain Management (SCM) elements.

This collaboration between actors of different specialties depends on exchanging and sharing adequate information on the product, related processes and business throughout the products' lifecycle. Thus, effective capture of information, and also its extraction, recording, exchange, sharing, and reuse become increasingly critical, especially when one considers the collaborators' *points of view*. All these lead companies to adopt new improved methodologies in managing the exchange and sharing of information, emphasizing closer collaboration between software systems that, up to now, have been operating independently.

This research is focusing on integration of product, process and organization modeling in a collaborative framework. This strategy improves the definition of the product development model, and the quality of the exchanged information by using *"multiple viewpoints"* connecting collaborators' features, and interests with appropriate parts of the product information-model.

In the next section of this paper, an overview of some interesting research on product information exchange topic, within the PLM and SCM, and on viewpoints architectures are presented and discussed. In Section 3, we describe our proposed collaborative framework based on Product/Process/Collaboration/Organization model (PPCO) and multiple viewpoints structure. An illustrated example of this framework is presented in Section 4. The last section summarizes the paper and gives an outlook for the future work.

## II. LITERATURE REVIEW

A product is made up of components, which are typically developed and managed by different departments or even outsourced. The components usually have their own definitions, which may not be identical (same interpretation) to the others companies' product definitions. Thus, to make a best collaboration between multidisciplinary actors, the organizations need to capture product development information and knowledge in the whole PLM processes. In the collaborative design case, Szykman et al. [2] [3] propose a knowledge-based approach for supporting engineering design, so-called "Design Repositories". Design repositories capture not only *what* is designed, but also *how* and *why* the product is designed. Design repositories enable to capture the product evolutionary nature of product knowledge and information throughout the design process. However, the system does not have provisions for modeling the product families and their evolution.

A recent effort for capturing knowledge to support knowledge-intensive design is the development of PLM technologies. Sudarsan et al. [4] present a conceptual framework for capturing product knowledge over the life of the product. The authors assert that two limitations are present in current PLM technologies. The first limitation deals with design changes and rationale throughout the product development process. This limitation is not within the scope of this work. However, the second limitation is that most PLM systems and enabling technologies are primarily focused on the form representation of the product. These efforts focus on the overarching scope of knowledge capture and reuse in product development; they offer an excellent basis on which to address the specific problem of knowledge reuse in engineering analysis. Thus, their proposed product information framework integrates only the design and assembly areas including product's family, tolerancy and product's evolution, and does not address collaboration issues which evolve heterogeneous systems.

Currently, Product Data Management (PDM) software approaches the functionality of design repositories and PLM systems most closely. PDM systems enable engineers to capture and share product information and meta-information across extended networks to design team members.

Gzara et al. [5] present an approach to build product models by reuse of patterns in product information systems engineering. The product information modeling is intended to represent and regroup data issued from various stages of product lifecycle. This information depends on the actor's activity, and is composed of diverse nature and complex elements (Simulation result, CAD model, bill of material, etc). So, this completes of the standard model of product to be able to treat all the product's abstraction levels: conceptual, generic and physical. All this information constitutes the informational flow elements of the various activities of the product development process.

In the Supply Chain (SC) area, some standards product data [6] [7] and information exchange systems have been proposed to support design practitioners in information management and exchanging data for concurrent engineering. Based on eXtensible Markup Language (XML) and Standard of the Exchange Product Data Model (STEP) [8], Burkett [6] proposed a new paradigm for product data exchange – namely the Product Data Mark-up Language (PDML) – to facilitate the integration and interoperability of the product development process. The PDML defines a set of application transaction sets for defining the data requirements to communities. In the same way, Lee et al. [9] present a logical model with object technology that can be converted to XML for supporting information exchange between a relational database and a knowledge repository. The authors attempt to propose dynamic data exchange schema based on the integration of a knowledge-based system with object technology and the XML standard.

On the multiple viewpoints and product views, we find Hoffmann [10] who proposes a mechanism for maintaining consistent product views in a distributed product information database. In his work, a single repository called a "master model" in which all-relevant product data resides was proposed for the integration of different product information domains, while the other views of the product must be updated to maintain consistency. This architecture builds on different applications distributed over different CAD/CAM systems. Thus Bronsvoort [11] proposed a multiple viewpoints feature modelling approach to allow a designer to focus on the information that is relevant for a particular product development phase. His system supports conceptual and assembly design, part detail design and part manufacturing planning by providing an own view on a product for each of these applications. Viewpoints as in [12] were introduced with the aim to ease the specification of complex software system. The main idea of this approach is to reduce the accessible aspects of the specification such that the developers only have to deal with those aspects of the system, which are important for the current task.

Some research [13] [14] [15] are focused on the collaborative systems in corporation with feature technology. Collaborative feature-based modelling systems are distributed; these systems provide concurrent working environment, synchronisation of modelling data for distributed users, and user interaction facilities.

The interests of presented works on the viewpoint area are two: 1) to define the integration of a professional point of view (e.g. design process) for all collaborators, and 2) to generate product views interpreted differently following the collaborator objective. The limitation of these studies is that none of them integrate the users' viewpoints about their interests within different product lifecycle phases. A multidisciplinary collaboration on the product lifecycle, including not only the geometric information but also other information (as manufacturing data, business data, etc.) directs us to integrate the multiple viewpoints user on the information itself.

In the next section, the proposed collaborative framework for the supply chain actors including PPCO and the multiple viewpoints approach are presented.

III. PROPOSED EXCHANGING AND SHARING INFORMATION FRAMEWORK

In intra and inter-organization Multidisciplinary Collaboration (MC), several collaborators exchange and share geometric data and other information on the product (structures, functions, business data, manufacturing data, bill of material, analysis data, marketing, resources, costs, etc.). To extract and exchange the right information following the actors' activities in the collaboration, and their focus on the product, we need to define a generic model on the different information linked to the product and process, and to develop some actor's viewpoints on the product (artifact or sub-artifact).

In Fig. 1 we present different links between the collaborators' activities (ellipse) and the domain (rectangle), based on seven profiles that regroup some of the SC collaborators: Design, Manufacturing, Supplying, Distribution, Maintenance, Client, and Partnering. These links are integrated in the viewpoints approach. (see Sect. 3.2)

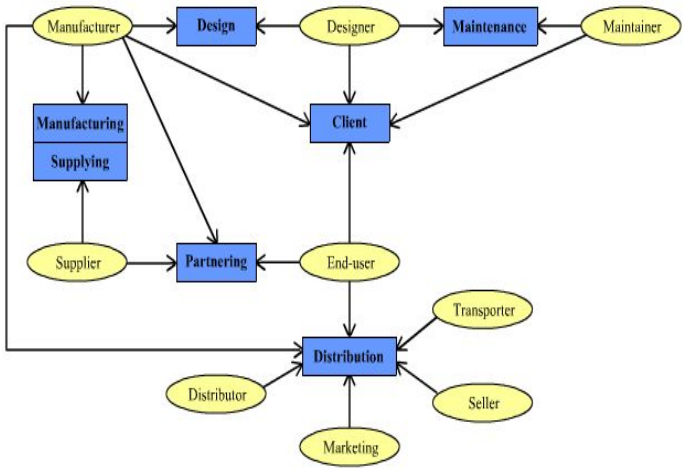

Figure 1. Activities and Domains links

In the following subsections, we describe our collaborative framework to exchange and share information among a supply chain context. This system is based on four layers (see Fig. 2):

- The *Product/Process/Collaboration/Organization* model (PPCO) is the core model of this architecture; PPCO defines product and process information, and the various collaboration methods for the organizations involved in the supply chain.
- *Viewpoint model*, which describes relationships between each actor and the comprehensive Product/Process model, defining each actor's "domain of interest" within the evolving product definition. This model allows each actor to retrieve automatically the adequate pieces of information required by his/her activity and focus.
- A layer defining the comprehensive *organization* (companies, functions, actors, etc.) and *collaboration relationships* between the actors within the SC.
- Based on the above relationships, the last layer proposes a *typology of exchanged messages*. A communication method, based on XML, is developed to support optimal exchange/sharing of information.

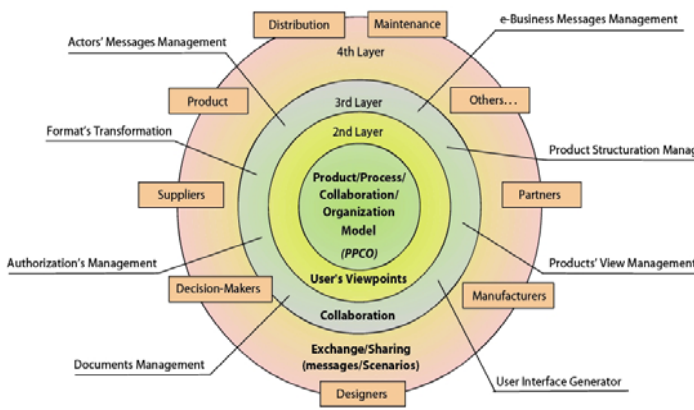

Figure 2. Proposal collaborative framework

In this paper, the two first layers are described in more details. To illustrate information structure, the information extraction and recording process. In the following subsections, we consider that each collaboration process has one or more collaboration-team with several actors (mono or multi-task) which will intervene jointly within collaboration. (see Fig. 3)

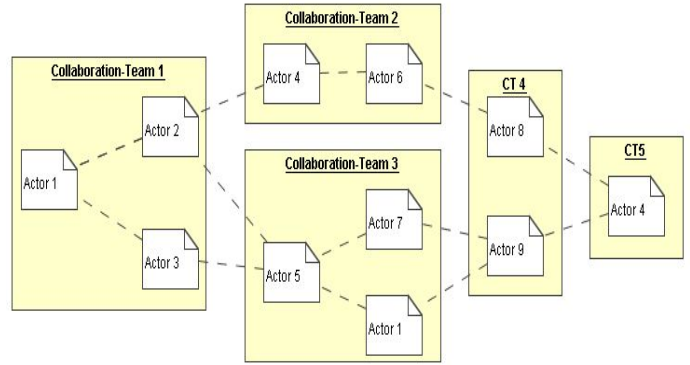

Figure 3. Example of collaboration's process

### A. The Product/Process/Collaboration/Organization Model

In multidisciplinary collaboration oriented PLM and/or SCM, we need to define a general core model which includes the information on the product (geometry, manufacturing data, etc.), on the process, on the collaboration, and on the structure of the supply-chain organization.

Some research on the product and process information definition, as [3], [4], [5], and [16], has motivated our research to regroup and define the core model for our architecture, including a generic model of SC collaboration to describe the different connection (exchange and share) between the collaborators' intra and inter-organizations.

Our first objective of creating the PPCO is to provide a base-level information model that is open, non-proprietary, generic, extensible, independent of any product development process and a generic collaboration in SC context, capable of capturing the full engineering and business context commonly shared in product development. So, the model is composed of four main parts interconnected directly or indirectly (e.g. using viewpoints) such, product, process, collaboration, and organization.

Fig. 4 illustrates the entities comprising the Product information. All entities are specializations of the abstract class **Flow_Entity**. **Technical_Entity**, **Control_Entity**, **Structural_Entity**, and **Statutory_Entity**, **Technical_Object**, **Relationship**, and **Statutory_Entity** are abstract classes. **Structural_Entity** describes the geometric definition and the interconnection between subcomponents, **Satutory_Entity** defines the technical object state (e.g. update_is_accepted, rejected, wait_validation). **Control_Entity** is used to define the framework interactions' exchange and constitutes the communication vectors between actors and the collaboration framework.

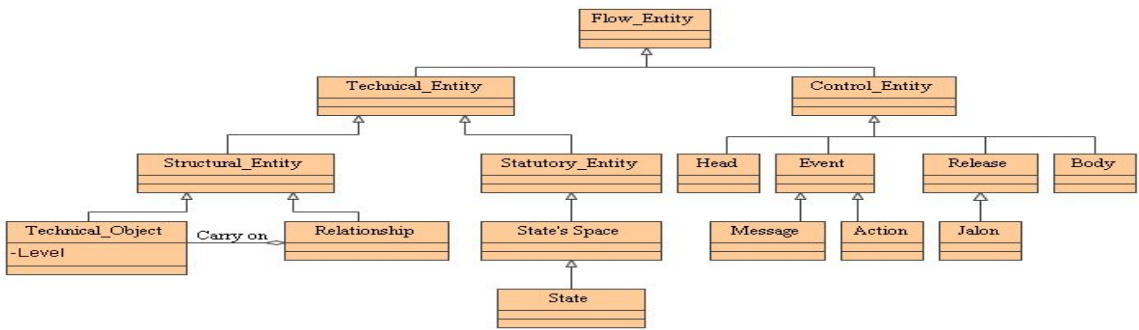

Figure 4. Technical and Control Product Information

Semantically, **Artifact** (Fig. 5) represents a distinct entity in a product, whether that entity is the entire product or one of its components. It is defined by three levels:

- *Abstraction level* which defines (a) Production process (manufacturing, support, and maintenance) relates to a physical artifact (b) Engineering process relates to the model of product according to which exemplars will be manufactured (product-type), and (c) Generic product.
- *Composition level* defines the assembly of the subcomponents to build the final product (Basis, Semi-finished, Finished)
- *Evolutionary level* defines the evolution of the product within the time and following customer's requirements, for this level, we include the indices' revision, state and time-constraints.

**Function** represents what the artifact is intended to do. The separate representation of Function renders the PPCO and its extensions capable of supporting functional reasoning in the absence of any information on the artifact's form. **Form** may be viewed as the proposed design solution to the problem specified by the function and consists of the artifact's **Geometry** (shape and structure may be synonymous in some contexts) and the **Material** it is composed of. **Behavior** represents how the artifact's form implements its function and is evaluated by a causal model [16].

**Relationship** is a relation which links technical object between them; the principal idea is to support a generic exploitation, in particular to seek constituent elements. **Constraint** gives the possibility to define conditions either on the links of subcomponents or on the application's domain. **Requirement** applies to some aspect of the function or form of the artifact. There are two specializations of **SetRelationship**: **Undirected_SR** groups objects into a set, while **Directed_SR** groups them into two subsets with different roles **Reference** links or cross-references entities. **Job_View** is a regrouping of links intended to facilitate the presentation of various Product information related to activity. For example, a view "Manufacturing" will gather the whole of the nomenclature's link, thus making it possible to an actor to explore the subcomponents.

In the process modeling, we were interested into the processes which contribute directly to the product development throughout its lifecycle: (1) Business process implements a business competence through the activities carried on within the framework of the principal functions related to the product (design, industrialization, manufacturing, maintenance). (2) Informational and/or Organizational processes which permit to manage the product information. We define **Process** as a set of activities (see Fig. 6), where the **Activity** itself represent the transformation actions of the input/output flows, it is described by (a) an **Objective** which define the process collaboration focus and the actor focus in the collaboration, and (b) some **Conditions** which depend on the activity and product context (post- and pre-conditions), and (c) **Resources** represent all entities which intervene on the flows' transformation (e.g. Actors, tools, machines, etc.).

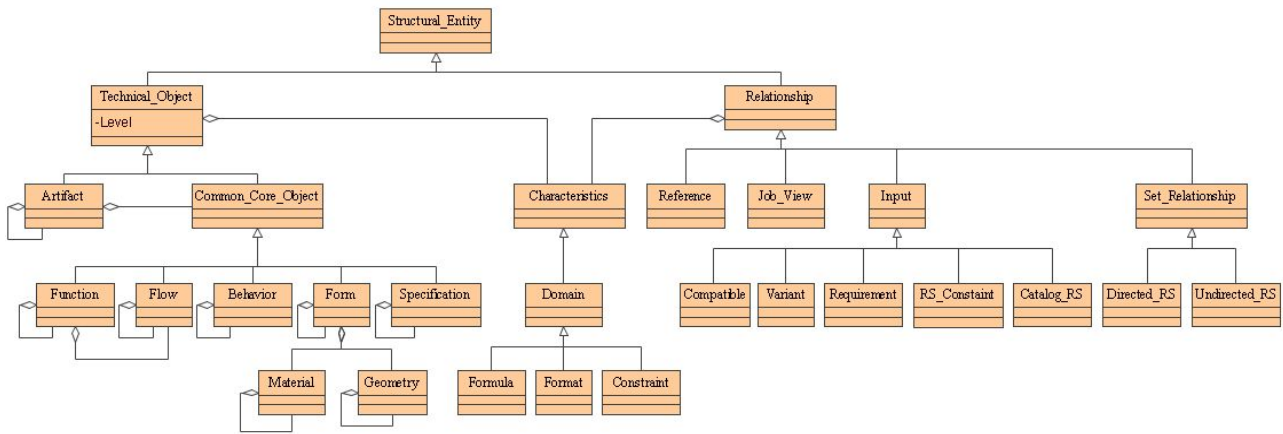

Figure 5. Technical Object Information and Relationship

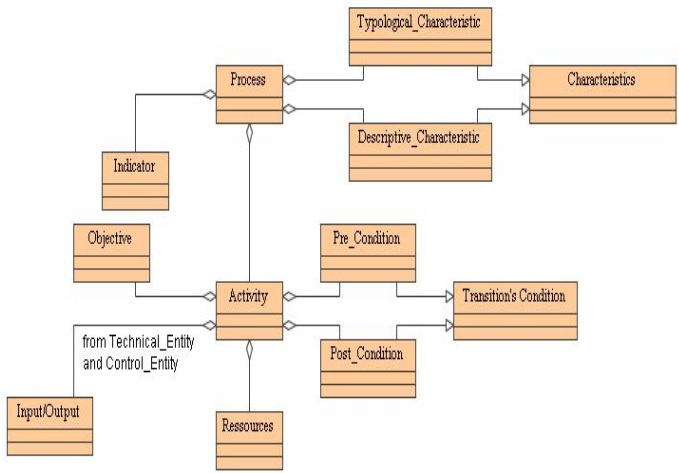

Figure 6. Process and Activity modeling

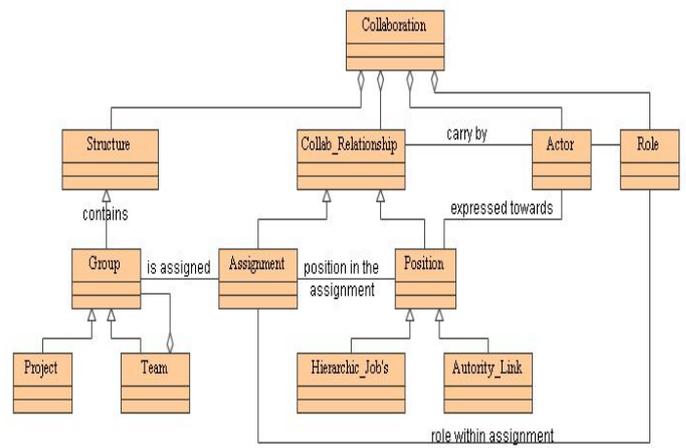

Figure 8. Collaboration modeling

After the definition of the product and the process modeling, we introduce the organization modeling (see Fig. 7), where any SC organization is described as a set of teams, and each team has a specific **Objective** and **Connector** with other teams or collaboration units. In this part, the customers define the team member and their respective **Role**. It would be illusory to modeling the interactions between actors without holding account of different collaborative-organization in which they cooperate, and where will be able to acquire their roles (see Fig. 8). In a collaborative structure, the sharing and exchange information relate always to a goal or an objective. It must be clearly identified by the partners so that own objectives can interact with the common objectives.

As-it was described above, at any level of process' collaboration the actor belongs to a team, and his/her activity is defined into an objective, and the conditions define the level of his competencies to facilitate the information extraction, and the activity's hierarchy in the team permit to optimize the interaction between actors and system.

Now, we gather these activities in several *families* which indicate the principal groupings and hierarchy of interesting information for the actors over the lifecycle (e.g. design analysis, shape design, product structure). Moreover, each family gathers several information *batches* (e.g. for the family of product structure, it consisted in the following batches: artifact, assembly, constraints, etc.).

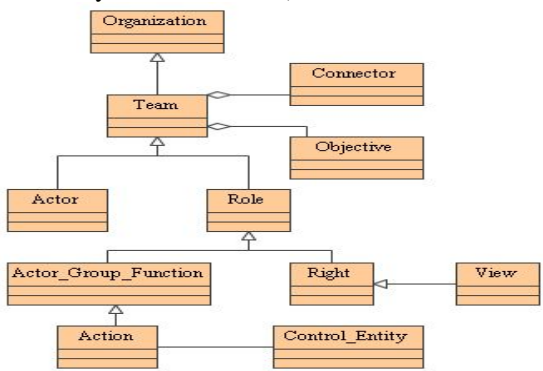

Figure 7. Organization modeling

To facilitate the comprehension of this decomposition, we introduce a "level" notation which determines its importance in the batch and accessibility to the user viewpoint (domain and focus). This level is defined by: (1) important information within a considered domain which includes it and can be used in another domain, (2) useful information in this domain and used by other domains, (3) superficial information within the domain.

After describing the core model (PPCO), we introduce in the next section the multi viewpoints approach to interconnect the PPCO and user needs.

### B. Multi-level viewpoints

The viewpoints framework provides an infrastructure for capturing and organizing product information extraction within multidisciplinary collaboration. Our notion of a viewpoint was of an object encapsulating cross-cutting and partial knowledge about activity, process, and domain of discourse, from the perspective of a particular collaborator, or collaboration-team, in the product development process.

Fundamentally, viewpoints organize product/software development knowledge based on separation of concerns. A viewpoint, expresses the concerns of a particular collaborator, such as a development participant or a representative of an area of concern captured by that viewpoint. Thus, a viewpoint may represent an area of concern within a project, a product, or a process, or may simply present a particular perspective expressed in a particular notation. The choice of dimensions of concern along which to create viewpoints may be the result of experience, the nature of the problem at hand, or simply organizational requirements.

The degree of importance and reuse of information changes from one to another collaborator following its objectives and activities sector on the product (see Fig. 1). The use of this information depends on different collaborators viewpoints, competencies, skills, responsibilities and interests on product information and product lifecycle phases. Several collaborators use the product information differently according to the specific requirements of their discipline. Starting from this section, we define the viewpoints as the collaborator's interest on the product, including his/her activity, process, domain, and focuses on the product; used to facilitate information extraction and collaboration within a Supply Chain context.

*1) Multi-level viewpoints description*

In a MC context, the human dimension is important and corresponds to the different participants in a product lifecycle, and a knowledge dimension corresponding to the experience, competence and situation of participants. Generally, we can say that a collaborator take a viewpoint according to his/her knowledge, domain of competence and his/her objective. Finch [17] speaks about a vocational viewpoint for a viewpoint used in a particular work activity.

So, in a MC context we notice that every user has one or more viewpoints following his/her activity in the different phases of product lifecycle process (see Sect. 4). A viewpoint is characterised by four objects (or concept types): **User**, **VP_Domain**, **VP_Relationship**, and **VP_Objective** (see Fig. 9 and Tab. I).

In Fig. 10, we consider two important parts in a viewpoint. First, the "VP_Objective" which constitutes the **focus**, second "VP_Domain" and "User" constitute the different **angles of view** on a same focus.

The concept types (user, domain, objective and relationship) have a different instantiation in the MC context to information restitution, but the information restitution is not always the same. (e.g. if the user has only one viewpoint on the product with any competence level, he does not need the Relationship attribute because he has only one viewpoint. If the user has only End-user viewpoint, he does not need the Domain and Relationship attributes.)

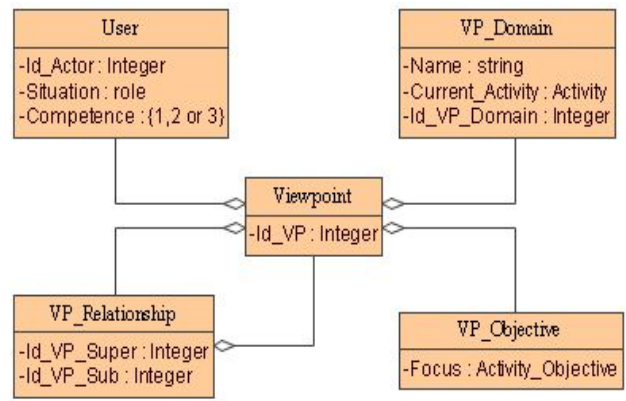

Figure 9.  Viewpoint definition

*2) Algorithm for filtering criteria*

In this section, we present one of the main information filtering algorithms (sub-processes of the main programs are not described in this paper). The filtering action is to find the appropriate information for a particular collaborator, according to his "viewpoints" on the product. However, the collaborators can carry out modifications, according to their rights, on restored information. For every modification, the user will submit his/her update which will be saved temporarily in the database (manually by modifying information one after one, or automatic modification by submitting of XML file), and annotated to all collaborators concerned by this modification. This update will take effect after approval of all concerned collaborators.

TABLE I. VIEWPOINTS CONCEPT TYPES

| | |
|---|---|
| **User** | is described by: <br>• *Name* of the collaborator or user. <br>• *Situation*, describes the role of the user through the collaboration, this role is defined by one of several values of situation-set = {Designer, Engineer, Supplier, Manufacturer, Distributor, Manager, Partner, Client, Other situation}. <br>• *Competence*, defines the level of competence for each user's activity, and it takes a value in {1, 2 or 3}. Every user must have only one activity with a level competence equal to 3. |
| **VP_Domain** | is defined by: <br>• Its *Name* <br>• Current *Activity* denotes the unique user's activity domain for each viewpoint (e.g. Electronics, Mechanics…), which is relied to the user's situation. |
| **VP_Relationship** | indicates the unique user's viewpoint which has the level competence equal to 3 on the same product. |
| **VP_Objective** | is described by: <br>• The *Focus* denotes the speciality in the user's activity (e.g. Topologic, Material, Thermal, Assemblage…) on a *Product*. |

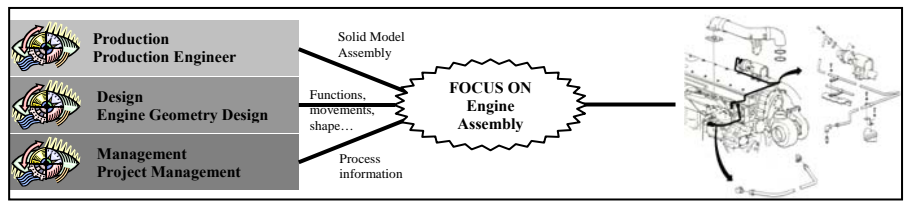

Figure 10. Example on a description of car-engine assembly

```
Program filtering_info_artifact (In: Artifact#, User#;
                                  Out: List_connexion_batch_info)
Begin
    // 1st step: Restitution of all user's viewpoints
    Restitution_list_viewpoint(In: User#; Out: List_vp_user);
    // 2nd step: Filtering of viewpoint on seek artifact
    Filtering_list_vp_artifact(In: List_vp_user, Artifact#;
                               Out: List_vp_user_prod);
    // 3rd step: Viewpoints classification by level of competence (decrease
      classification)
    Classification_vp(In: List_vp_user_prod; Out: List_vp_class);

    For i = 1 to size(List_vp_class) Do
    Begin
    // 4th step: Restitution of information batches and the expertise level for each
      viewpoint in using the focus, activity and level competence
    Restitution_list_connexion_level(In: List_vp_class;
                                     Out: List_connexion_level_vp);
    // 5th step: Update of the list of information batches and its level using
    If (i>1) Then
            Optimize_list_connexion_level(In: List_connexion_level_vp;
                                          Out: List_lev_vp_prebious);
    Else
            List_con_lev_vp_previous = List_connexion_level_vp;
    End_If
    End_For

    List_connexion_batch_info = List_connexion_level_vp;
End
```

## C. Collaboration relationship

Collaboration tools are one of the most effective technologies in today's competitive business environment, particularly, in the product customization, development, manufacturing, supply and services. According to the nature of the collaboration itself, the collaboration can be classified into five levels, which are communicative, collective, cooperative, coordinated, and concerted. In this level, we try to describe a generic collaboration's relationship within intra- and inter-organization. One part of this level is modeled on the PPCO; we integrate a product collaborative space by connecting different collaboration-teams, and tracing of information evolution from the inputs of collaboration process to its outputs.

## D. Typology of exchanged message

Based on the above levels, we propose a typology of exchanged messages to communicate and optimize information exchange and share it between actors. To define the messages' structure, we studied and compared several business frameworks. We noticed that it exist three categories of frameworks: (1) Cross-industry frameworks provide cross-industry vocabularies but are limited to the rough process approach and are not greatly concerned with messaging (e.g. xCBL, OAGIS). (2) Specific-industry frameworks give industry-specific vocabularies. However, their main contribution is in business processes, as papiNet (www.papinet.org) and RosettaNet (www.rosettanet.org) which provide a comprehensive description of business processes in a particular industry by applying the detailed process approach. (3) Process-centric frameworks provide no vocabularies but focus on business processes taking the generic process approach (e.g. ebXML, BPML). In our case, we integrate the OAGIS, papiNet and RosettaNet messages to create a standard XML messages exchange following the business process.

## IV. APPLICATION

To illustrate this methodology, we present an example about a Piston of automotive engine (Fig. 11) (Object#: 381009). We present also two viewpoints of same actor/user (Fig. 12).

The user "Georges" (with identifier: 18936) has two viewpoints in the system on the product. To retrieve the appropriate information for its focus on the product, we proceed as follows (see the steps in the last algorithm):

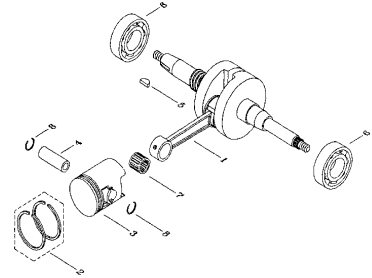

Figure 11. Piston of car-engine (Object#: 381009)

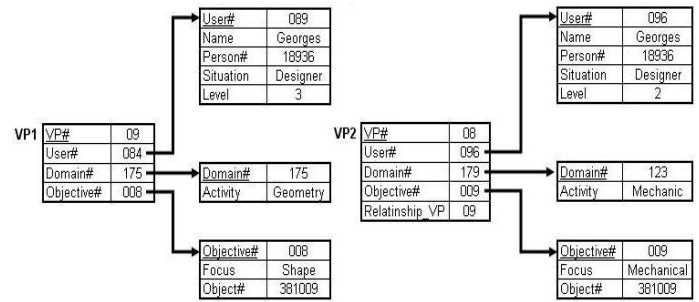

Figure 12. Two viewpoints of Georges in the System

*Step 1:* Retrieve all viewpoint of user with User#= 18936

*Step 2:* Filtering and classification viewpoints for the Piston of engine (Product#: 381009)

TABLE II. THE LIST OF ALL VIEWPOINTS OF USER "GEORGES"

| VP_List_Prod | VP# | 09 | 08 |
|---|---|---|---|
| | Situation | Designer | Designer |
| | Level | 3 | 2 |
| | Activity | Geometry | Mechanic |
| | Focus | Shape | Mechanical |
| | Product# | 381009 | 381009 |

*Step 3:* Filtering of information's batches of each viewpoint

```
PV# = 09 (VP1)
Batch (Level):
    • Artifact (1) : Standard Information about the product
    • Function (2) : Different function of the final-product and sub-artifact
    • Behavior (2) : Different behaviors of the final-product and sub-artifact
              in relation with their respective functions
    • Flows (2) : Different flows of the final-product functions
    • Geometry-Form (1) : All information about the detailed-geometry
              with CAD model
    • Sub-Artifact (2) : Different information about the second level of
              direct-component assembly
    • Assembly (2) : In relation with Sub-Artifact
    • Constraints (1): All constraints of the product (design, assemblage…)
    • Requirements (2) : Requirements about the product and different
              phases of its lifecycle
    • Group (1) : All information about participants of collaboration-team
```

```
PV# = 08 (VP2)
Batch (Level):
   • Mechanic (1) : All information about the mechanic application of the
                    product (see activity of VP09 and VP08)
   • Artifact (2) : Standard Information about the product
   • Function (2) : Different function of the final-product and sub-artifact
   • Behavior (2) : Different behaviors of the final-product and sub-artifact
                    in relation with their respective functions
   • Flows (3) : Different flows of the final-product functions
   • Geometry-Form (2) : Different information about the geometry with
                         CAD model
   • Sub-Artifact (3) : Different information about the first level of direct-
                        component assembly
   • Assembly (3) : In relation with Sub-Artifact
   • Constraints (1) : All constraints design of the product
   • Requirements (3) : Requirements about the product and different
                        phases of its lifecycle
   • Group (1) : All information about the participants in the group
   • Thermal (2)
   • Material (2)
   • Manufacturing (2)
```

Following the level's batch definition, the system regroups the batches with high-level hierarchy, and retrieves the information which is more adequate to the user following its focuses on the product, and its activities on the collaboration.

## V. CONCLUSION & DISCUSSION

In this paper, we presented our contribution to improve the exchange and sharing collaborative framework. This framework is based on four layers and instead to capture, exchange, share and reuse information within supply chain context. The integration of collaboration and sc-organization, within product and process information model, will permit to widen its scope of application than PDM systems. The viewpoints approach was integrated to capture the actors' interest on their collaboration over product lifecycle. Of course, this framework is intended to help managing the SC information and has no effects on the collaboration strategies or protocols. The viewpoint mechanism can be undeniable contribution for optimizing the product information exchange. With the integration of multi-viewpoints approach, the system particularly ensures:

- *The integration and consolidation* of information coming from various sources, and its filtering, transformation and adaptation to various viewpoints needs.
- The viewpoints permit to *optimize the time of information's restitution* and its contents while retrieval only the important information for the user.
- *The access* to permit information in real-time for collaboration actors (throughout management process).
- *The information traceability* permits to understand the information evolution from one to another state (using index, versions, and authentications of information).
- A true remote team work, it will make it possible to exchange "product" information and ideas. Effectiveness of creating an exchange and sharing environment of technical information.

As perspectives, we consider to improve definition of the collaboration relationship by integrating some standard scenarios of collaboration. Also, we must to define and integrate other types of viewpoints, especially process' viewpoint on the information and the relationship between user's and process' viewpoints to improve the information update.